\documentclass{article}
\pdfoutput=1%For ARXIV

\usepackage{amsmath,amssymb,amsfonts,amstext}
\usepackage{graphicx}
\usepackage{subfigure}
\usepackage{textcomp,nicefrac}
\usepackage{siunitx}
\usepackage{textgreek}
\usepackage{stfloats}
\usepackage{mhchem}
\usepackage{authblk}
\usepackage{xcolor}
\usepackage[a4paper, total={6.5in, 9.5in}]{geometry}

\usepackage[style=ieee,minnames=3,maxnames=3,backend=biber]{biblatex}
\DeclareSourcemap{
	\maps[datatype=bibtex]{
		\map{
			\step[fieldset=editor, null]    % 清除 editor 字段
			\step[fieldset=issn, null]
			\step[fieldset=url, null]      % 清除 issn 字段
			\step[fieldset=date, null]      % 清除 issn 字段
			\step[fieldset=note, null]      % 清除 issn 字段
			\step[fieldset=month, null]
		}
	}
}

\usepackage{hyperref}
\hypersetup{
	hypertex=true,
	colorlinks=true,
	linkcolor=blue,
	anchorcolor=blue,
	citecolor=blue
}

\usepackage{lineno}

\graphicspath{{figure/}}

\newcommand\nuclide[2]{$ ^{\text{#2}}\text{#1} $}
\newcommand\D[0]{\mathrm{\Delta}}
\newcommand\sss[1]{\,\si{#1}}
\newcommand\dbd[0]{0νββ}

\begin{document}
	
	\title{Pre-study of a \ce{Li2MoO4} based bolometer  for \nuclide{Mo}{100} neutrinoless double beta decay experiment in China}

        % authors
        \author[a,b]{Deyong Duan}
        \author[a,b]{Mingxuan Xue\thanks{ Correspondence to: 96, JinZhai Road, Baohe District, Hefei, Anhui, 230026, P.R. China. \\ \qquad E-mail address: xuemx@ustc.edu.cn (M. Xue).}}
        \author[a,b]{Kangkang Zhao}
        \author[a,b]{Taiyuan Liu}
        \author[a,b]{Haiping Peng}
        \author[c]{Jiaxuan Cao}
        \author[c]{Long Ma}
        \author[d]{Liang Chen}
        \author[d]{Hui Yuan}
        \author[a,b]{Qing Lin}
        \author[a,b]{Zizong Xu}
        \author[a,b]{Xiaolian Wang}
        
        \affil[a]{Department of Modern Physics, University of Science and Technology of China, Hefei, 230026, China}
        \affil[b]{ State Key Laboratory of Particle Detection and Electronics,  University of Science and Technology of China, Hefei, 230026, China}
        \affil[c]{Key Laboratory of Nuclear Physics and Ion-Beam Application (MOE), Institute of Modern physics, Fudan University, Shanghai, 200433, China}
        \affil[d]{Shanghai Institute of Ceramics, Chinese Academy of Science, Crystal Centre, Shanghai, 200050, China}
        
	\date{}
	
	\maketitle

	\begin{abstract}
		
  The cryogenic phonon scintillating bolometer is a promising and extremely attractive option to search for the nuclide neutrinoless double beta decay. In this paper, a pre-study of bolometer based on \ce{Li2MoO4}  (LMO) crystal is presented,
  %A project of searching for \nuclide{Mo}{100} neutrinoless double beta decay with \ce{Li2MoO4}(LMO) cryogenic phonon scintillating bolometers at China Jinping Underground Laboratory is proposed. In this paper, the pre-study of a LMO bolometer is presented. 
  in which the properties of the LMO crystal at the low temperature, including scintillation characteristics and specific heat, are investigated in detail. %as the key parameters for bolometer design, have been detailed investigated experimentally.  
  The excitation spectrum and light yield are measured from the room temperature down to 10 K, and 
  heat capacity is measured down to temperature of $O(200)$ mK.
  Furthermore, 
  %ranging from 10 K to room temperature; and the heat capacity measured down to $O(200)$ mK demonstrates that the result is consistent with the prediction of Debye model. Consequently, 
  a $\sss{(2~cm)}^3$ cubic LMO based bolometer is manufactured and tested at ultra-low mK-level temperature in a ground-above cryostat platform, and a good energy resolution is achieved.
  The studies laid a foundation to manufacture the bolometer detector in China and conduct neutrinoless double beta decay research at the China Jinping Underground Laboratory.
  \\
  
  %running in an above-ground cryostat at ultra-low mK-level temperature. The energy resolution as FWHM has achieved at 24.6 keV@511 keV.
  
		\noindent Keyworks: Neutrinoless double beta decay, Lithium molybdate, Bolometer, Scintillation property, Low-temperature heat capacity.
		
	\end{abstract}

	%\linenumbers
	
\section{Introduction}
	
    The study of Standard Model (SM)-forbidden process, neutrinoless double beta decay (\dbd), is a frontier research in the field of particle and nuclear physics. 
    If it occurs, it directly proves the Majorana nature of neutrinos~\cite{0vDBD-5}.
    Meanwhile, the nuclide half-life time of \dbd ~($T^\text{\dbd}_{1/2} $)  yields a strong constraining on the 
    % If \dbd, the Standard Model (SM)-forbidden process, were observed, it would directly confirm the Majorana nature of neutrinos\cite{0vDBD-5}. 
    % In addition, the sensitivity of $ T^\text{\dbd}_{1/2} $(the half-life time of \dbd) yields strong restrictions on 
    neutrino absolute mass~\cite{0vDBD-6}, which is extremely valuable to answer the absolute mass scale and mass hierarchy of neutrinos~\cite{0vDBD-8}.  
	
\begin{figure}[htbp]
		\centering
		\includegraphics[height=0.35\textwidth]{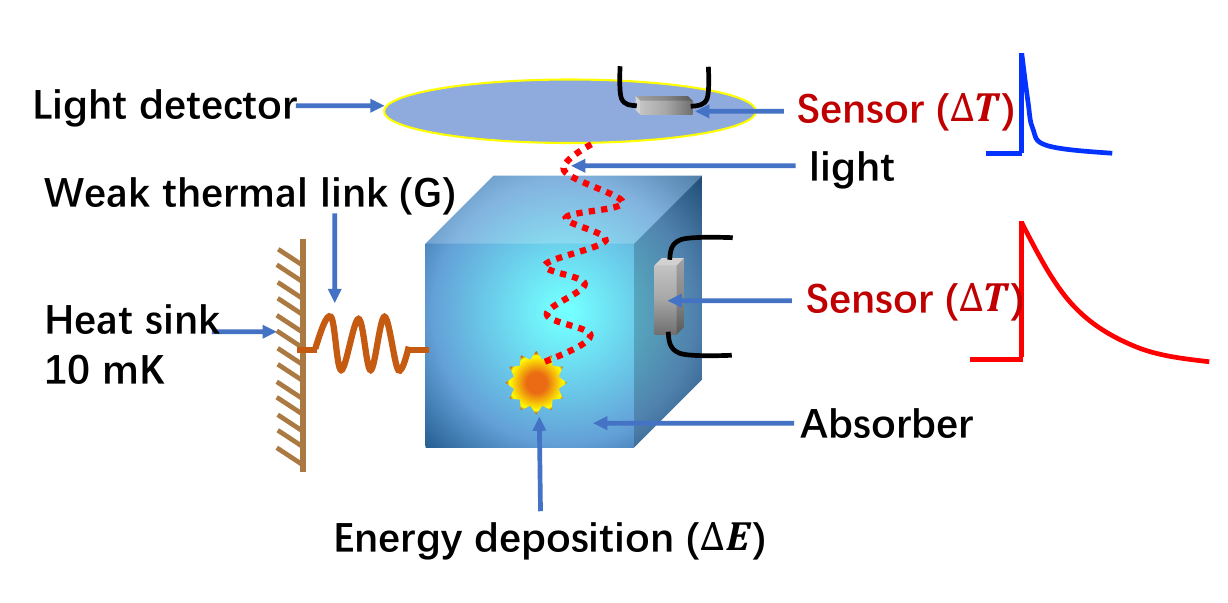}
		\caption{Schematic diagram of the cryogenic phonon-scintillating bolometer. There are five main components: ({\romannumeral1}) absorber, a scintillation crystal, where the particles interact and deposit energy; ({\romannumeral2}) light detector, which collects the florescent photons and then convert them into heat;
			({\romannumeral3})  sensors, which measure the temperature increase of the absorber and the light detector; ({\romannumeral4}) heat sink and ({\romannumeral5}) weak thermal coupling, which provide stable low temperature and energy release.  }
		\label{01bolo}
    \end{figure}

    Up to now, several different detection techniques are applied to hunt for \dbd, and the cryogenic phonon scintillating bolometer is one of the most competitive ones. %are becoming one of the most competitive detection techniques in the search for \dbd. 
    The detector principle for the cryogenic phonon scintillating bolometer is shown in Fig.~\ref{01bolo}, and details can be found in Ref.~\cite{gironi2010scintillating}.    
    %Figure~\ref{01bolo} shows the detector principle. 
    At present, the molybdate crystals are attractive to serve as absorbers of bolometer~\cite{LMO01}, since \nuclide{Mo}{100} is a \dbd\ candidate nuclide with high Q-value (3034~keV) and relatively high natural isotope abundance (9.82\%)~\cite{01Qvalue}.
    The molybdate crystals are favored by CUPID-Mo experiment at Modane in France~\cite{cupid}, AMoRE experiment at Yangyang in South Korea~\cite{amore}, CROSS experiment at Canfranc in Spain~\cite{cross} and etc. 
    A world largest lithium molybdate (\ce{Li2MoO4}, LMO) crystal array, the CUPID experiment, with more than 1000 LMO crystals at LNGS in Italy is on the R\&D phase, and is expected to commission in the near future. 
    Meanwhile, there are many other R\&D projects in progress with various molybdate crystals~\cite{CMO,ZMO,PMO}. 
    Particularly, a proposal aiming to explore the \dbd\ of \nuclide{Mo}{100} nuclide with LMO crystal is promoting at China Jinping Underground Laboratory (CJPL)~\cite{cjpl},
    % there is a proposal at China Jinping Underground Laboratory (CJPL)~\cite{cjpl}, which aims to explore the \dbd\ of \nuclide{Mo}{100} with lithium molybdate (LMO). The CJPL 
    which has its own natural advantages, covered by a 2400 m rock layer and is an unique platform for low background experiments. 

	To explore the feasibility of cryogenic phonon scintillating bolometer based on LMO crystal, a pre-study is carried out in China. 
    In the rest of this paper,
    %of searching for \nuclide{Mo}{100} \dbd\ using LMO light-heat dual readout bolometers, the rest of this paper is arranged as follows. 
    Sections~\ref{sec2} and~\ref{sec3} present the studies of the scintillation properties and the specific heat of LMO crystal at low-temperature, respectively.    
    Section~\ref{sec4} shows the assembly process of a bolometer based on LMO crystal. 
    Section~\ref{sec5} presents the test process and performance of bolometers at mK-level temperature.
    %assembly process with a LMO crystal, and shows its performance at mK-level temperature on a Platform for Cryogenic Detector R\&D, located at an above-ground laboratory at University of Science and Technology of China(USTC), aimed at bolometer R\&D for \dbd\ search. Finally, 
    Section~\ref{sec6} shows the conclusion and discusses future prospects.

\section{The scintillation properties of LMO }

	\label{sec2}

    For a cryogenic phonon scintillating bolometer with light-heat dual readout system, the scintillation characteristics of the absorber at low temperature are critical. %for the scintillating bolometer performance with light-heat dual readout system. 
    The light yield of crystal (absorber) directly determines the 
    %On the one hand, the light yield, which describes the amount of fluorescent light for unit energy deposition in the absorber, directly affects the 
    discrimination capability between \textalpha\ and \textbeta/\textgamma ~as well as the light-heat signal coincidence efficiency of the bolometer.  
    %On the other hand, 
    Meanwhile the emission spectrum of the crystal is critical for the design of light detector where 
    %in determining the material selection of the light detector, since the 
    the semiconductor used for the light detector is required to have  band gap energy smaller than the energy of the single emitted photon. 
    Additionally, the emission spectrum is strongly related with the thickness of optical %file
    film coated on the light detector, which is applied to enhance the  light transmittance~\cite{ARC}. 
    
    %In addition, in order to enhance the transmittance of the light detector, it is necessary to coat it with optical film. The thickness($ d $) needs to satisfy Equ. \ref{12}, where $ n $ is the refractive index of the optical film and $\lambda$ is the wavelength of the fluorescent light\cite{ARC}. Therefore, the emission spectrum, especially the wavelength information, is helpful to fabricate the light detector with better performance.
    %\begin{equation}
    %    nd = \frac{1 + 2k}{4} \lambda, \quad ( k = 0,1,2 ... )
    %    \label{12}
    %\end{equation}

\begin{figure*}[htbp]
		\centering
		\includegraphics[height=0.35\textwidth]{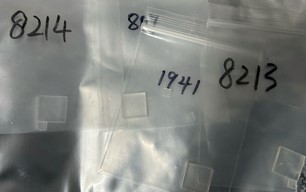}
		\caption{LMO crystals grown by SICCAS. The slices are $ 10\sss{mm} \times 10\sss{mm} \times 2 \sss{mm}$ in size. }
		\label{02setup}
	\end{figure*}

    At present, the absolute light yield and the emission spectrum of crystal at mK-level temperature cannot be measured directly due to the absent of test platform.
    But their tendencies varying from room temperature to the low temperature are the alternative approach to understand these characteristics. 
    The scintillation properties at the various temperature from 300~K down to 10~K are studied 
    %in our current experimental setup, so the trend from room temperature to low temperature as a substitutional method is hired. 
    %The scintillation properties of LMO are studied from 300 K to 10 K 
    using a HORIBA Fluorolg Tau-3 spectrofluorometer~\cite{02eq} for the LMO crystal sample with a size  $ 10\sss{mm} \times 10\sss{mm} \times 2\sss{mm}$, as shown in Fig.~\ref{02setup}, where the crystal samples are produced by SICCAS (Shanghai Institute of Ceramics, Chinese Academy of Science). 
	
\begin{figure*}[htbp]
		\centering
		\includegraphics[width=0.4\textwidth]{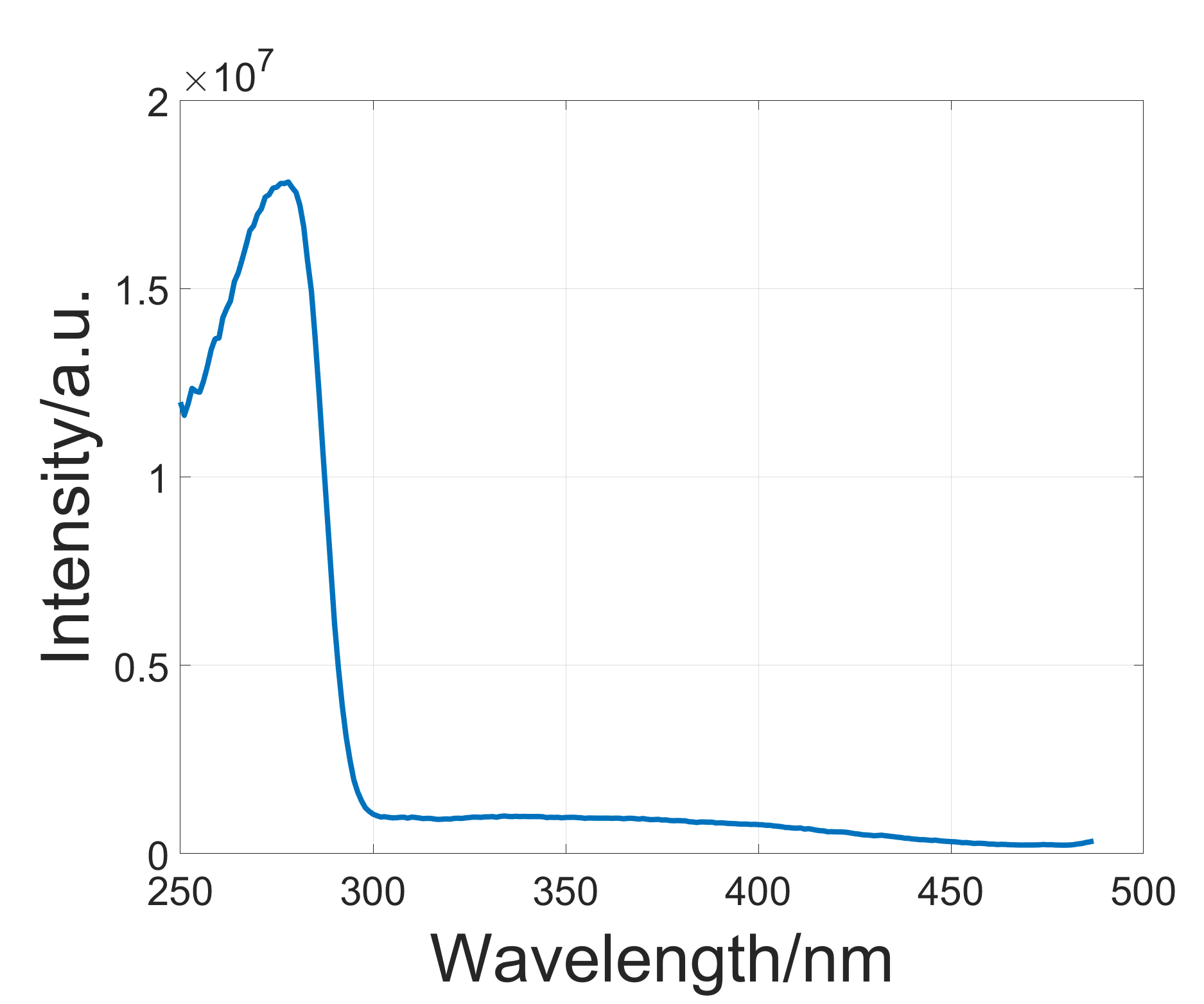}
		\caption{The excitation spectrum of LMO at 10 K, peaked at $\sim$280 nm. }
		\label{02lmoex}
	\end{figure*}

\begin{figure*}[htbp]
		\centering
		
		\subfigure[]{\label{02lmoem}
			\includegraphics[width=0.35\textwidth]{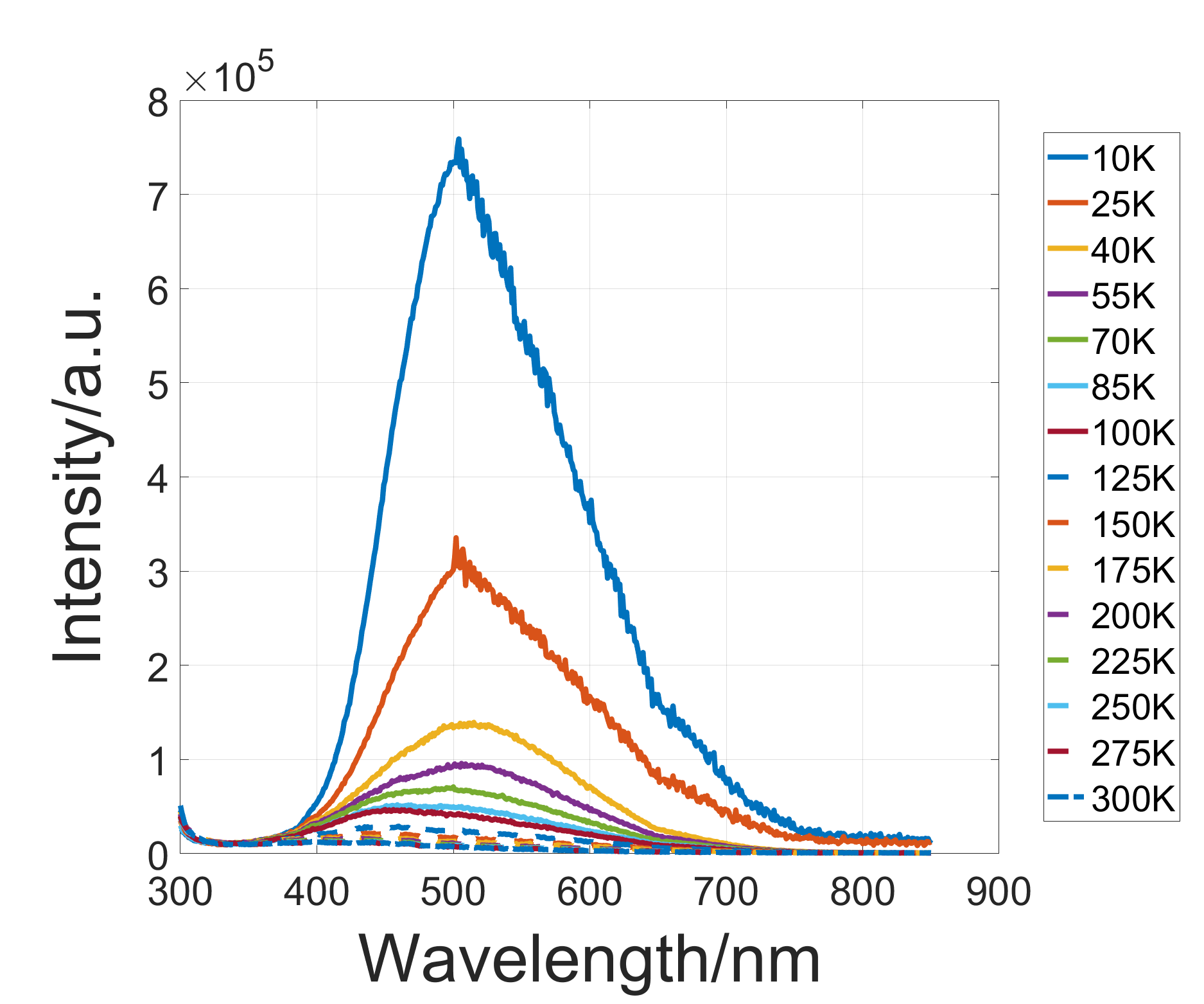}}
		\subfigure[]{\label{02lmotem}
			\includegraphics[width=0.35\textwidth]{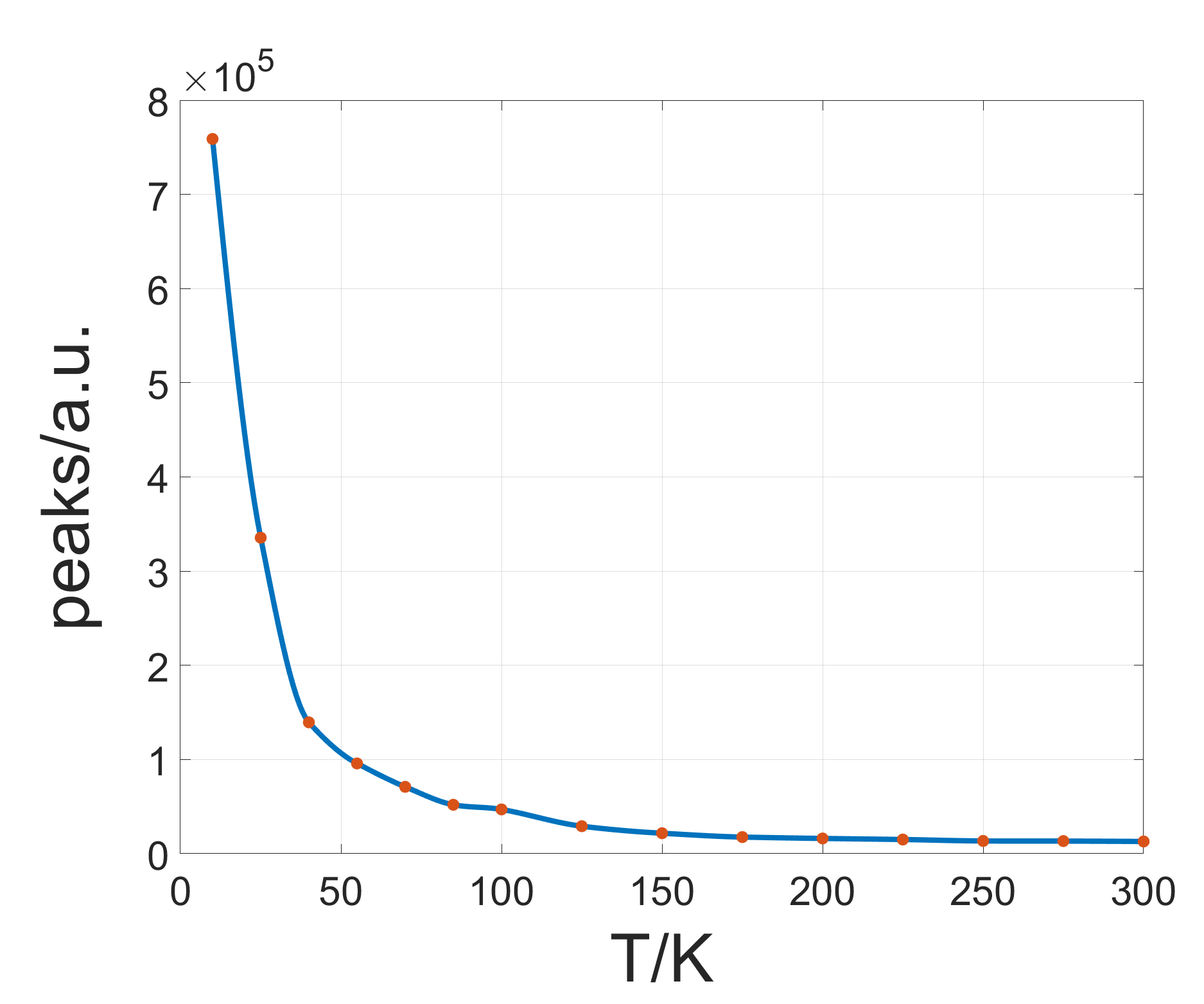}} 
		
		\caption{(a) A laser-induced emission spectrum of a LMO crystal at different temperatures, peaked at 510 nm. (b) Temperature dependence of the luminescence intensity of a LMO crystal. }
		
	\end{figure*}

   To obtain the excitation spectrum of LMO crystal, we performed a laser scan with wavelength varying from 250 to 480~nm at a temperature of 10~K. As shown in Fig.~\ref{02lmoex}, the excitation spectrum of LMO crystal peaks at a  wavelength of $\sim$280 nm. 
   Then the tendencies of emission light spectrum and yields varying from 10 to 300~K are studied by exciting the LMO crystal with a laser beam of a wavelength of 280~nm.
   As shown in Fig.~\ref{02lmoem}, the emission spectrum distributes broadly with a wavelength from 350 to 850~nm. The peak of emission spectrum shifts higher slowly when the temperature descenting and at a peak wavelength of 510~nm for a temperature at 10~K. Figure~\ref{02lmotem} shows the temperature dependence of light yields of the LMO crystal, the light yields descends when the temperature increases, especially when the temperature below 50~K, the light yields varies rapidly. The fluorescent lights become very faint at a room temperature. The observed varying tendencies of emission light spectrum and yields are as the expectation of molybdate crystals.
  
   %Figure~\ref{02lmoex} presents the excitation spectrum of LMO with laser wavelength from 250 nm to 480 nm at a temperature of 10 K, where its peaks at wavelength $\sim$280 nm. 
   %Meanwhile, the LMO crystal excited with a laser beam of 280-nm wavelength exhibits broad emission band that peaking at 510 nm as shown in Fig. \ref{02lmoem}. At room temperature, the emission light of LMO crystal becomes very faint. The lower the temperature, the higher the light yield. LMO has fluorescent lights at low temperatures such as 10 K. Figure \ref{02lmotem} shows that the measured temperature dependence of the relative light yield of the LMO crystal. The result follows the rule of the molybdate crystal family, where the light yield at low temperature is higher than that at room temperature.

	\section{The specific heat capacity of LMO at low-temperature}
	\label{sec3}

    The heat capacity ($C$) of the absorber is a key parameter for the cryogenic scintillating bolometer, and is directly related with its sensitivity and energy resolution.
    %According to the basic principle of the cryogenic phonon detector, the heat capacity ($C$) of the detector absorber is a key parameter. 
    For a given energy deposition ($\D E$) in an absorber, the smaller heat capacity, the higher bolometer sensitivity ($\D T = \D E/C $, where $\D T$ is a slight temperature change of the absorber). 
    To optimize the detector performance, it is necessary to measure the cryogenic heat capacity of the absorber precisely.
    As the inorganic insulating scintillation crystals, the heat capacity of LMO crystal at very low temperature ($T \ll \Theta_D$, where $\Theta_D$ is the Debye temperature) can be described with the Debye model as~\cite{03Debye} :
    \begin{equation}
        C(T) = \dfrac{12}{5} r N_{\text{core}} k_B \pi^4  \left(     \dfrac{T}{\Theta_D}     \right)^3
    \end{equation}
    where $N_{\text{core}}$ is the total number of molecules or unit cells in the core absorber, $r$ is the number of atoms in each unit cell, $k_B$ is the Boltzmann constant. %, and $\Theta_D$ is the Debye temperature of the absorber material. 

\begin{figure*}[htbp]
		\centering
		\includegraphics[height=0.3\textwidth]{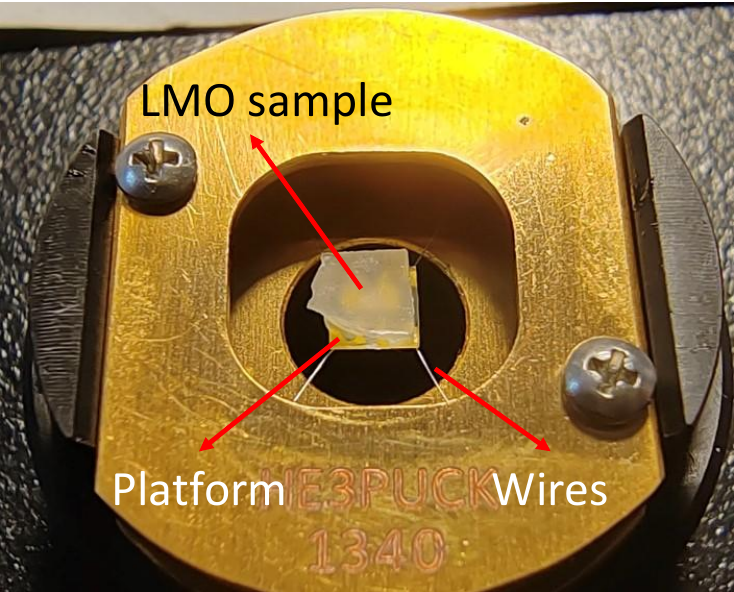}
		\caption{LMO crystal sample with a mass of 14.3 mg glued on the testing platform. The wires are for supporting the sample. }
		\label{03sample}
	\end{figure*}

\begin{figure*}[htbp]
		\centering
		\includegraphics[width=0.48\textwidth]{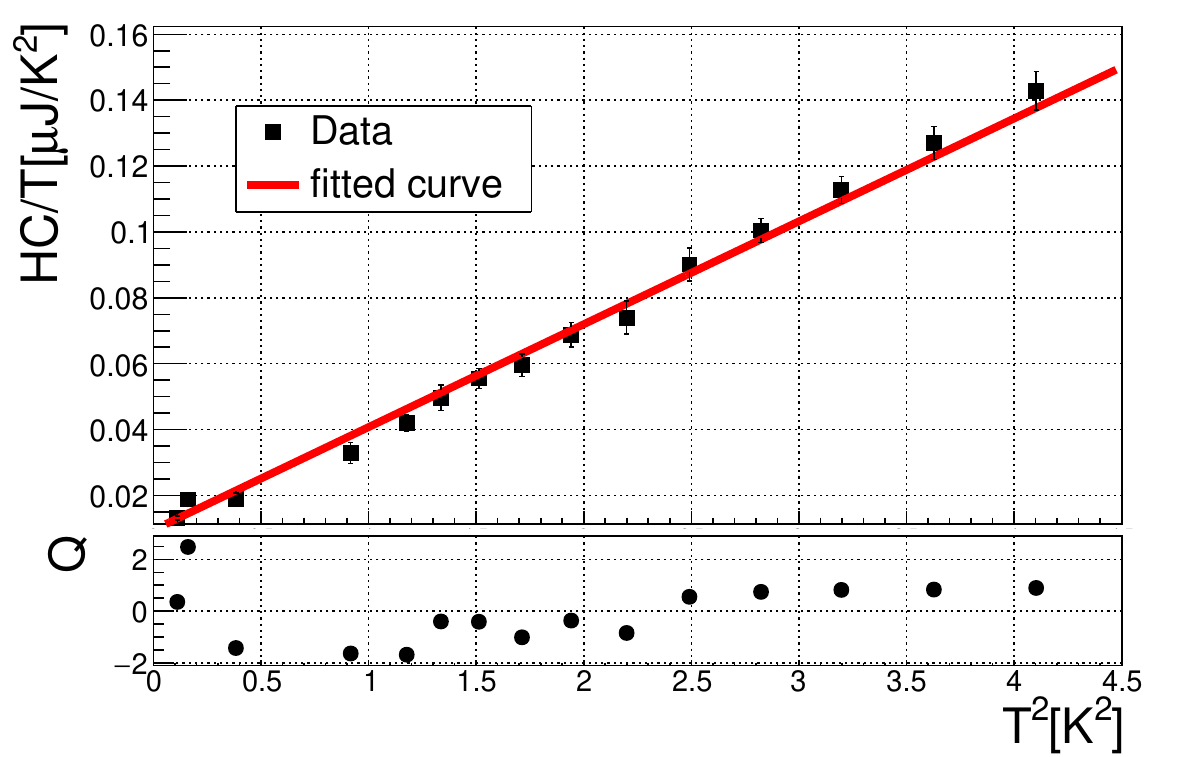}
        \caption{ The ratio $C/T$ as a linear function of $T^2$. The error bars of  heat capacity are determined by PPMS model fitting software. The Q value is defined as the residuals divided by the error of data to represent the deviation between data and the fitting result.}
		\label{03data}
	\end{figure*}

    The heat capacity of a LMO crystal sample with a mass of 14.3~mg and flat surface is studied within a temperature varying from 200~mK to 2~K by a Quantum Design Physical Properties Measurement System (PPMS)~\cite{03PPMS} based on a relaxation technique~\cite{03relax}.
    The detailed processes of measurement and calculation can be found in Ref.~\cite{PWO}.
    Figure~\ref{03sample} illustrates the LMO crystal sample glued on the testing platform.
    %In this study, a LMO crystal sample with a mass of 14.3~mg and flat surface sits on the testing platform, as shown in Fig.~\ref{03sample}, and the heat capacity is measured by varying the temperature.
    The ratio of measured heat capacity to the temperature ($C/T$) is presented in Fig.~\ref{03data}. %where a good linear dependency is observed.
    The distribution is fitted with a linear function, 
    %data plots showing the ratio $C/T$ as a linear function of $T^2$ can be found in Fig. \ref{03data}.
    %Red line in Fig. \ref{03data} shows the linear fitting result as Equ. \ref{HC-T},
	\begin{equation}
		\label{HC-T}
		C/T = \gamma + \beta\cdot T^2, 
	\end{equation}
   and the resultant fit curves is shown with red line in  Fig.~\ref{03data}.

	The slope of the linear function from the fit can be used to obtain the heat capacity and the Debye Temperature ($\Theta_D$), which is $ \Theta_D(\text{LMO})=(330\pm3) \sss{K}$. Despite the temperature region difference, this result aligns with the conclusion of another measurement~\cite{03LMOHC}.
	In general, at these low temperature, the heat capacity of the diamagnetic insulating crystal is proportional to $T^3$. The intercept term is rejected since it may come from the systematic uncertainty of the testing platform, which is also observed in the previous study~\cite{PWO}.

	%\newpage
	
	%\section{Development of a 2-cm cubic LMO bolometer and detector performance }
        \section{LMO based bolometer desigh and assembly}
	\label{sec4}
\begin{figure}[htbp]
		\centering
		\subfigure[]{
			\label{04bolo}
			\includegraphics[height=0.35\textwidth]{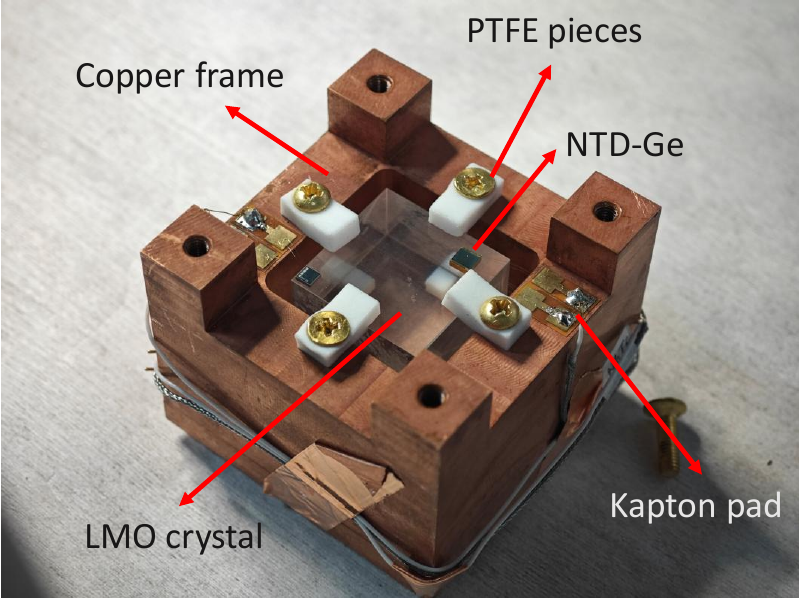}}
		\subfigure[]{
			\label{04spring}	
			\includegraphics[height=0.35\textwidth]{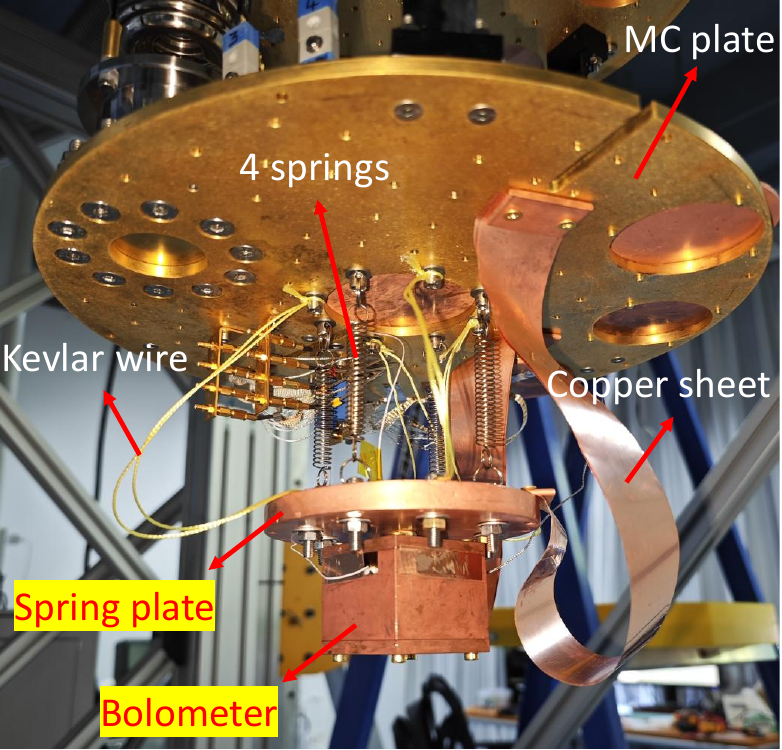}}
		\caption{(a) The %$\sss{(2cm)^2}$ 
  $\sss{(2cm)^3}$cubic LMO bolometer is assembled inside a thick copper frame. %PTFE pieces are employed to fix the crystal and play a role as the thermal link. An NTD-Ge semiconductor thermometer is glued onto the surface of the LMO crystal to measure thermal fluctuations induced by incident particle interactions, applying the Kapton pad as the electronic connection bridge. 
                 (b) The LMO bolometer is mounted on the spring floating plate in a cryostat at USTC. }
		\label{}
	\end{figure}

	Based on the above properties studies, a cubic bolometer detector with heat channel readout only is design and assembly, where the absorber is the above same LMO crystal produced by SICCAS.
    To suppress the pile-up effects from radioactive background, the LMO crystal is relatively small,  a size of $ 2\sss{cm} \times 2\sss{cm} \times 2\sss{cm}$ with a mass of 21.7~g. The surfaces of crystal are polished to optical grade.
 %the low-temperature properties testing of LMO, a pre-study of a LMO bolometer with only heat channel naturally begins. The LMO crystal produced by SICCAS has a mass of 21.7 g and a size of $ 2\sss{cm} \times 2\sss{cm} \times 2\sss{cm}$, and all the surfaces are polished to optical grade.	
 %	\subsection{LMO bolometer design and assembly }	
	The LMO crystal is housed inside a copper frame with 1.5 cm thick, which is a bit thick by considering no specialized shielding for the test platform during the operation. 
    %Considering that there are no specialized shielding materials such as lead and copper in this run, the copper frame is slightly thick. 
    PTFE pieces serve  as the weak thermal link, fixed with brass screws. 
    All components have been thoroughly cleaned. 
    A semiconductor temperature sensor NTD-Ge~\cite{ntd}, which has features of large dynamic range, high sensitivity and relatively simple readout electronics, is adopted for the heat signal readout.
    %The heat signal readout of LMO adopts an NTD-Ge sensor \cite{ntd}, which is a semiconductor temperature sensor with large dynamic range, high sensitivity, and relatively simple readout electronics. 
    The dot matrix method based on Araldite epoxy is applied for the coupling between LMO crystal and NTD-Ge sensor.
    %The coupling between LMO and NTD-Ge applies the dot matrix method based on Araldite epoxy. 
    A Kapton pad is glued on the edge of the cooper frame for transferring electrical signals. 
    Four gold wires with 25~\textmu m diameter  are used for the electrical connection between NTD-Ge and the Kapton pad by wire bonding, then the twisted superconducting NbTi wires are soldered from the Kapton pad to the Micro-D joint on the cryostat.   
    The design and assembly of the LMO crystal based bolometer are shown in Fig.~\ref{04bolo}.

\begin{figure*}[!h ]
		\centering
		\subfigure[]{
			\label{04ustcdr}	
			\includegraphics[width=0.45\textwidth]{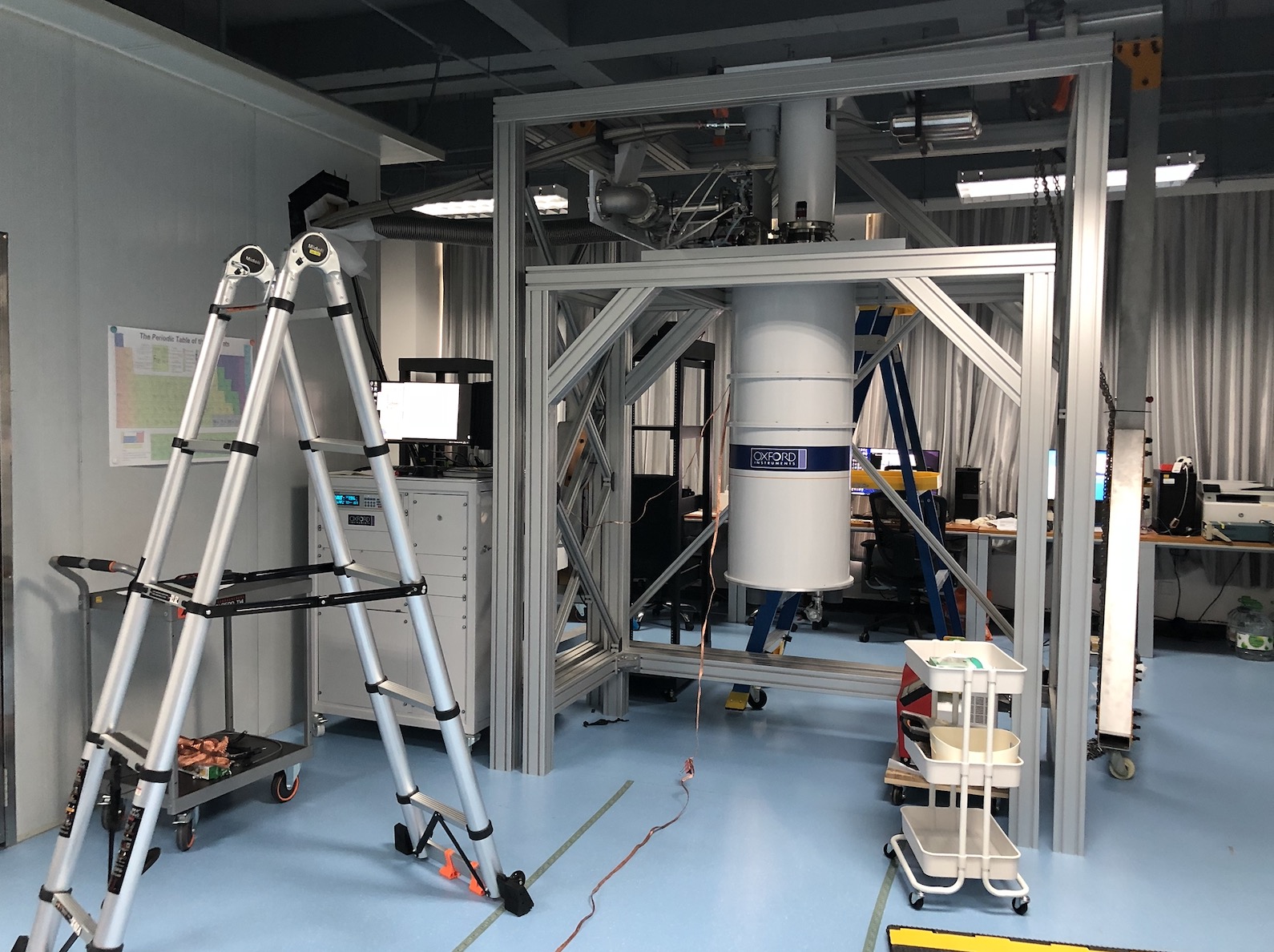}}
		\subfigure[]{
			\label{04water}
			\includegraphics[width=0.45\textwidth]{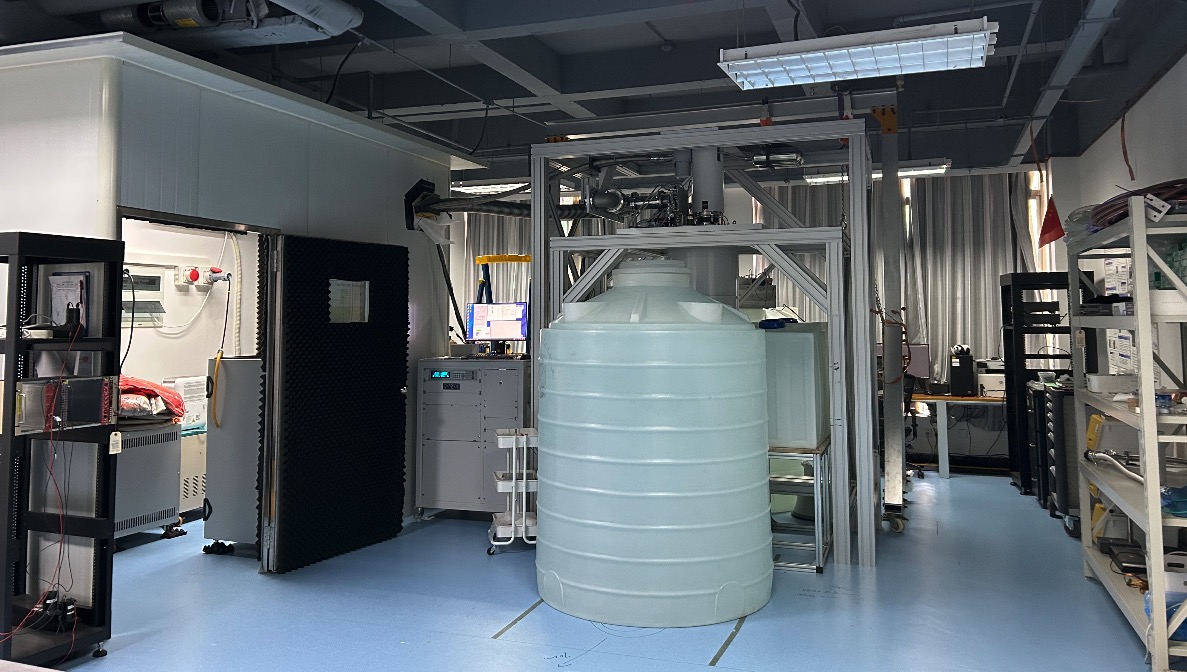}}
		\caption{The ground-above cryostat platform used to test the LMO based bolometer located at USTC. (a) The Oxford Triton-500 dilution refrigerator. (b) Running with simple water shielding to reduce the surrounding environment backgrounds. }
		\label{}
	\end{figure*}

    The assembled LMO bolometer is mounted on a spring plate in a cryostat platform, as shown in Fig.~\ref{04spring}.
    The cryostate platform is a dilution refrigerator of Oxford Triton-500 located in a ground-above laboratory at University of Science and Technology of China (USTC), as shown in Fig.~\ref{04ustcdr}.
    Previous studies have demonstrated that the vibration of Pulse Tube is one of the main noise backgrounds for detector performance. 
    Therefore, four simple stainless-steel springs are connected to a floating plate from the Mixing Chamber (MC) plate to suppress the mechanical vibration. 
    To prevent damage caused by sudden breakage of the springs at mK-level temperature, Kevlar wires are integrated in the system for safety protection. 
    Four high-purity oxygen-free copper sheets serve as thermal conductors between the two plates. 
    In addition, a thermometer is placed on the spring plate as well to monitor the real-time temperature during cooling-down phases. 

   \section{Detector operation and data analysis}	
   \label{sec5}
	%\subsection{Running in a cryostat}
    %as shown in Fig. \ref{04spring}. The dilution refrigerator is an Oxford Triton-500 (shown in Fig.  \ref{04ustcdr}) at the Platform for Cryogenic Detector R\&D, located in an above-ground laboratory at USTC. It is used for bolometer R\&D in the search for \dbd. 
    The cryostat platform is operated stably with a temperature of 20~mK at the spring plate and 5~mK at the MC plate in the cryostat during the test. 
    %In this low temperature test of the LMO bolometer, when the temperature stabilizes, the temperature of spring plate is 20 mK and that of the MC plate is around 5 mK. \textbf{
    Since no dedicate radioactive shielding is existed in the platform, a simple water shielding, as shown in  Fig.~\ref{04water}, is implemented.  
    The working point is selected when NTD-Ge has its maximum resistance.
    %due to the relatively large pileup from the radioactive background.
    %The working point during the operation only considers the maximum amplitude of the detector only, where the NTD-Ge is at its maximum resistance value. 
    %the detector is in an above-ground cryostat and there is no effective dedicated radioactive shielding (but with simple water shielding, as shown in Fig. \ref{04water}), the accumulation of numerous pile-up events from the environmental radioactivity makes it impossible to adopt the conventional method for optimal working point selection. 
    The resistance value of NTD-Ge is around 200~kΩ, corresponding to a temperature of LMO crystal of 28~mK.
    The signal of NTD-Ge is amplified, and fed into a data acquisition system without trigger.
    The data stream eventually are recorded by a 16-bit ADC (NI-6218) with a 10-kHz sampling frequency. 
    For the energy calibration during the operation, a radioactive source, \nuclide{Na}{22} (with typical gamma energies of 511~keV and 1274~keV), is set up outside the cryostat.
    
    %platform.

\begin{figure}[htbp]
		\centering
		\includegraphics[width=0.45\textwidth]{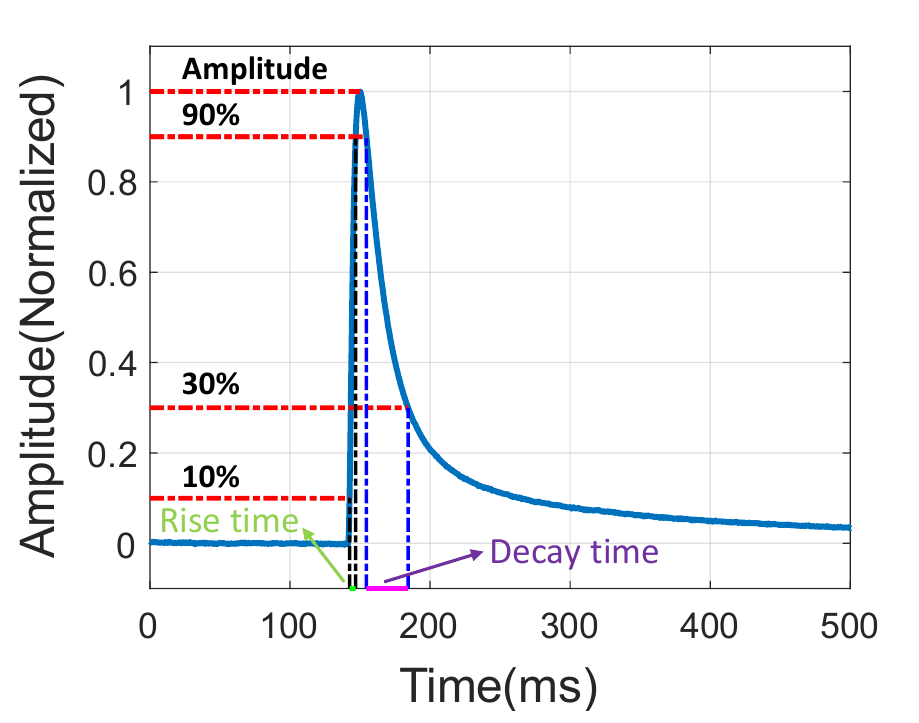}
		\caption{The LMO bolometer heat pulse in a 0.5-second time window. The typical decay time is 29.8 ms and the rise time is 4.3 ms.}
		\label{04pulse}
	\end{figure}

    %For the 10 hours of data analyzed in this study, the resistance value of NTD-Ge is around 200 kΩ, while the temperature of the LMO is about 28 mK. The pulses read by NTD-Ge are amplified, fed into a data acquisition system, and recorded by a 16-bit ADC (NI-6218) with a 10-kHz sampling frequency. Besides that, there is a radioactive source, \nuclide{Na}{22} (with gamma lines at 511 keV and 1274 keV), outside the cryostat for energy calibration during the measurement. 
	%\subsection{Detector operation and data analysis}
    The data taking is continued for 10 hours. Figure~\ref{04pulse} illustrates a typical 0.5-second time window displaying the temporal evolution of LMO heat pulses. 
    The collected triggerless data stream is processed by a Python-based visualization software tool~\cite{zhao}, which exploits a matched filter, the Gatti-Manfredi Optimal Filter, to maximize the ratio of signal-to-noise (S/N), and to extract the pulse-shape parameters, including amplitude, decay time, rise time, baseline and etc. 
    
\begin{figure*}[htbp]
		\centering
		\subfigure[]{
			\label{04cal}	
			\includegraphics[height=0.33\textwidth]{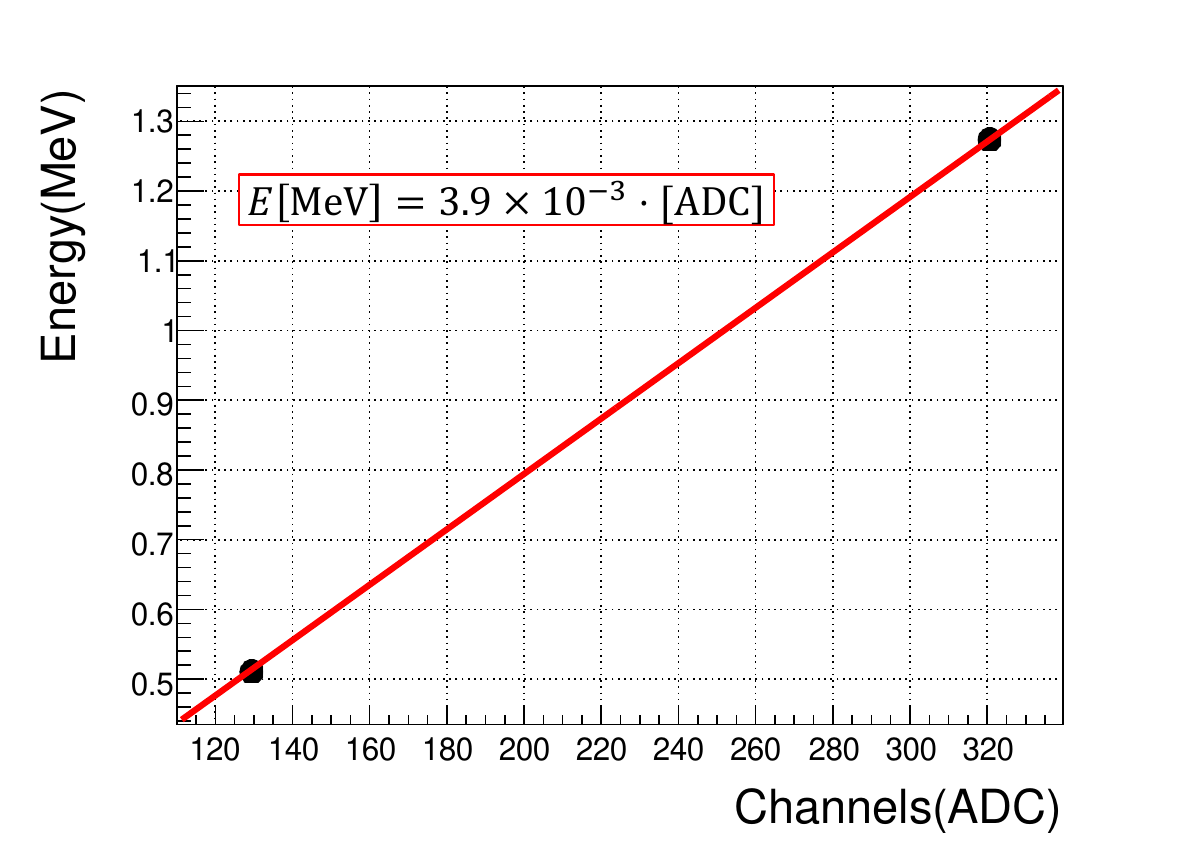}}
		\subfigure[]{
			\label{04eng}
			\includegraphics[height=0.33\textwidth]{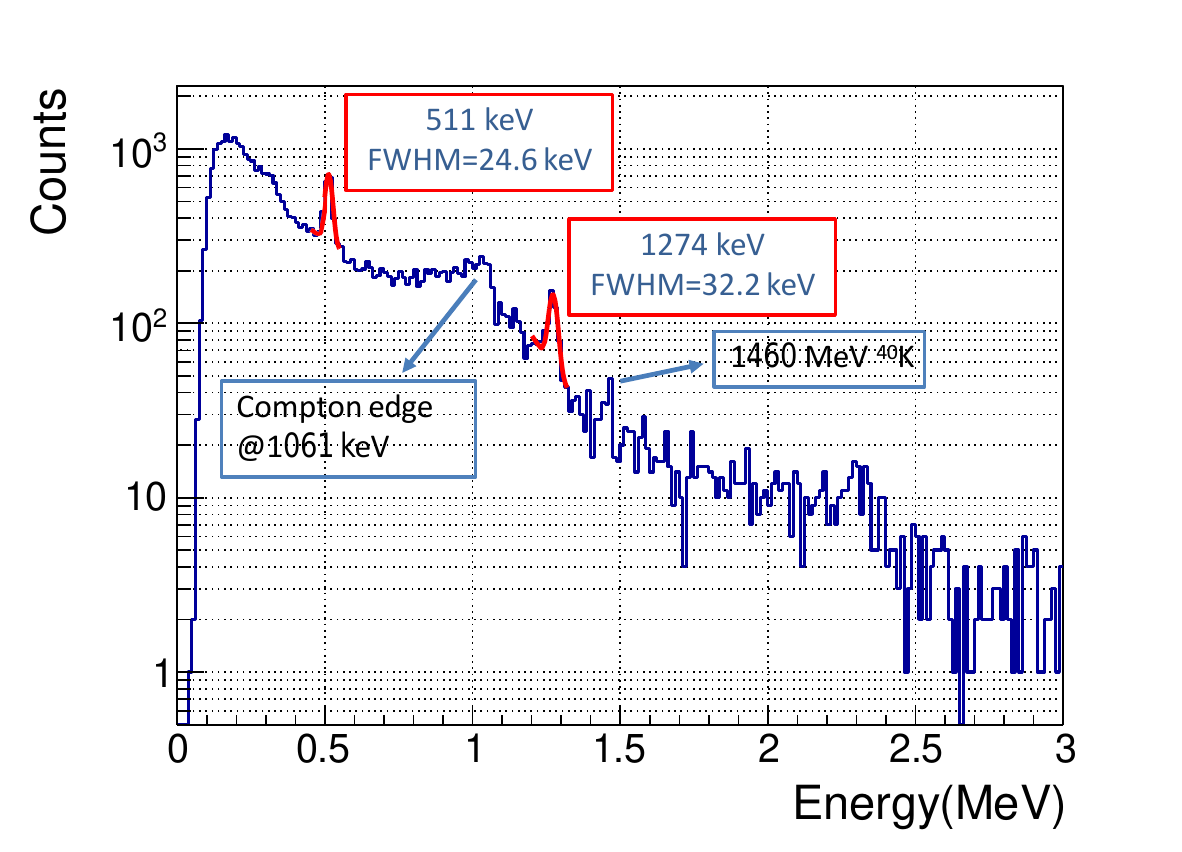}}
		\caption{(a) Energy calibration of the LMO based bolometer using $\gamma$ events of 511~keV and 1274~keV from \nuclide{Na}{22}. (b) The Energy spectrum of the LMO based bolometer for 10 hours data taking. Beside the clear peaks at 511 and 1274~keV (\nuclide{Na}{22}), the structure of the Compton edge at 1061~keV from 1274~keV photons can also be observed, the peak at 1460~keV is from \nuclide{K}{40}.}
		\label{04res}
	\end{figure*}

    The energy calibration is carried out based on the two typical $\gamma$ energy, 511~keV and 1274~keV from the \nuclide{Na}{22} calibration source.
    %For energy calibration of the LMO bolometer, two typical γ lines with relatively high intensity are used, \nuclide{Na}{22} (511 keV) and \nuclide{Na}{22} (1274 keV). 
    The fits are performed on the ADC distribution in the two corresponding $\gamma$ signal regions, individually, where the signal is described with a Gaussian functions, and background with a polynomial function. 
    The resultant means of Gaussian function correspond to the real energy of $\gamma$, as shown in Fig.~\ref{04cal}.
    The calibration constants are obtained by performing a fit on the really energy as a function of ADC bin with a linear function, as shown in  Fig.~\ref{04cal}, too.
    The distribution of energy after the calibration of data is shown in Fig.~\ref{04eng}, where beside the two known calibration $\gamma$ signal located at the 511 and 1274~keV, there is clear bump peaking around 1061~keV corresponding the compton edge of  $\gamma$ from \nuclide{Na}{22} with energy 1274~keV. 
    There is also a small peak at 1460~keV, corresponding to the typical $\gamma$ from nuclide \nuclide{K}{40}.
    To evaluate the performance of bolometer, the fits are carried out on the energy distribution within the range of two calibration $\gamma$ of  \nuclide{Na}{22} source, individually, and yields the FWHM of energy are 24.6 keV@511 keV and 32.2 keV@1274 keV.     
    %The mean value of the Gaussian functions are regarded as the specific ADC where the full energy peaks locate. 
    %A linear fitting is performed on energy as a function of corresponding ADC, as shown in Fig.~\ref{04cal}. 
    %An important indicator for evaluating bolometer performance is energy resolution. The fitting is made with ROOT v6.32.02 applying the log likelihood method as shown in Fig. \ref{04eng}. The FWHM of typical γ peaks are evaluated to be 24.6 keV@511 keV and 32.2 keV@1274 keV. 
    Though the obtained energy resolution are not competitive with the world leading values, but it proves the bolometer detector works well. Many optimized avenues, including more effective radioactive shielding, more stable and lower temperature operation, longer time data acquisition, improvement of analysis software, enhancement of LMO crystal quality and so on are expected in the future refinements. %These refinements will further enhance the bolometer's sensitivity and stability. 
	
	\section{Conclusion}
	\label{sec6}

 The cryogenic phonon scintillating bolometer based on the molybdate crystals as absorber is promising and extremely attractive to search for the nuclide neutrinoless double beta decay (\dbd). In this paper, the scintillation properties at a temperature range 10$\sim$300~K and heat capacity at a temperature range 200~mK$\sim$2~K are studied for the LMO crystal produced by SICCAS.
 The emission spectrum of LMO crystal distributes broadly with a wavelength from 350 to 850~nm, and peaks at 510~nm at a temperature of 10~K.
 The light yield increases as the temperature descents, especially varies rapidly when the temperature below 50~K. 
 The heat capacity of LMO crystal follows the Debye model with a Debye Temperature $\Theta_D = (330\pm3)$~K in the range of 200~mK$\sim$2~K.
 Based on the above studies, a $\sss{(2cm)^3}$ cubic LMO cryogenic phonon scintillating bolometer with the heat-only readout is designed and manufactured, and the performance is tested on a ground-above cryostat platform at USTC at a temperature of 28~mK.
 This work yields the energy resolution FWHM of 24.6 keV@511 keV, and we have got the direction for the potential space to improve the detector performance in the near further.
 % the potential to improve the performance are promoted and can be applied in the future.
 % Furthermore, the fluorecent light properties measured show that LMO crystal is a suitable absorber for a heat-scintillation bolometer, a bolometer with light-heat double readout using LMO is going to be fabricaited.
 Furthermore, given the fluorecent light properties of LMO crystal, a light-heat double readout is going to be developed as an advanced approach to improve the particle discrimination capability.
  %ranging from 10 K to room temperature; and the heat capacity measured down to $O(200)$ mK demonstrates that the result is consistent with the prediction of Debye model. Consequently, 
  The studies in this paper laid a foundation to manufacture the bolometer detector in China, and to conduct neutrinoless double beta decay research at the China Jinping Underground Laboratory.

    %LMO scintillation crystals are promising absorbers for cryogenic phonon detectors in the search for \nuclide{Mo}{100} \dbd. 
    %It is meaningful to study their low-temperature properties, which are key parameters for the LMO bolometer R\&D design. 
    %We have systematically investigated the low-temperature characteristics of the LMO crystal produced by SICCAS, focusing on fluorescence yield and heat capacity. 
    %Regarding to the heat capacity, the result follows the Debye model, $C\propto T^3$, where the Debye Temperature $\Theta_D(\text{LMO})$ is $(330\pm3)$ K in the range of 200 mK – 2 K. 
    %Building on these research, a pre-study of a 2-cm cubic LMO bolometer with heat-only readout is successfully tested in an above-ground cryostat. 
    %The energy resolution has obtained at 24.6 keV@511 keV with 10-hour running at 28 mK without specific radioactive shielding. While further optimization is required to enhance the detector performance, these preliminary results demonstrate the feasibility of LMO bolometer at USTC. In summary, the release of these results can support and promote the proposed project of bolometer demonstration experiment at China Jinping Underground Laboratory.

	\section*{Acknowledgment}
	
	This work is supported by National Key R\&D Program of China (2023YFA1607203), National Natural Science Foundation of China (12475190, 12141505, 11875310), the Fundamental Research Funds for the Central Universities, China (WK2360000015), the State Key Laboratory of Particle Detection and Electronics, China (SKLPDE-ZZ-202403). We extend our appreciation to Dr. Zhi Zhao and Dr. Jiyin Zhao at the Physical and Chemical Science Experiment Center, University of Science and Technology of China, for their expert assistance in the measurement of scintillation properties and specific heat, respectively.

	\newpage
	
	\printbibliography
	
\end{document}